\documentstyle[12pt]{article}
\begin{document}
\begin{center}
{\Large What is the manifestation of a ``quasar'' at
$z > 10^{10}$ ? }
\end{center}

\begin{center}
{\large   V. K. Dubrovich}
\end{center}

\begin{center}
Special Astrophysical Observatory RAS\\
          St. Petersburg 196140, Russia\\
          dubr@MD1381.spb.edu
\end{center}

\begin{abstract}

The process of forming an image of a cosmological point 
source (CPS)
in condition of high optical depth is considered accounting
for all types of interactions.
It is shown that the energy conservation law causes the
size of this image which is keeping constant over
all redshifts of the CPSs.
This effect must be taken into account for 
the consideration of the angular power
spectrum of the CMBR. 
In particular, distant point sources and small scale
fluctuations which  
were damping before recombination will contribute their 
energy in
the region of angular scale $\theta_0 \approx 20'$.

\end{abstract}

{\bf 1}.  One of the main processes in the evolution of matter 
in the Early Universe
is a process of increasing and dissipation any density 
fluctuations which
may be adiabatic fluctuations produced by inflation, 
primordial black holes, inhomogeneous of distribution of matter 
and antimatter, etc. Many of
them are relatively small in size and rapidly evolve. 
We consider those of
them which become smaller then the horizon before 
the hydrogen recombination. It seems that
they does not produce any observable effects. 
A lot of articles claim their invisibility.
The simplest argument is that the optical depth by the
Thomson scattering $\tau \gg 1$. 
It becomes larger than unity at redshift $z \ge 1100$ and
rapidly increase at higher $z$ :
\begin{equation}
   \tau = n_e\sigma_T c t \approx 10^{-2}\,(1+z)\: ,
\end{equation}
where $n_e$ is the concentration of free electrons 
(and positrons), $\sigma_T$ the
cross section of the Thomson scattering, $c$ the
speed of light, and $t$ the cosmological time.

{\bf 2}.  But in our case there is a different situation. 
It is not a point source 
which is obscured by very thick cloud. 
Our source is surrounded by
homogeneous matter with a very small free path for photons. 
This situations
leads to a very slow diffusion of photon from the point 
where it was emitted.
It means, that photon cannot move far away from 
the initial point. In a
medium with a constant $n_e$ 
it can spread for the time interval $t$
in a bubble with the radius $R$ 
\begin{equation}
   R \approx ct/\tau^{1/2}\: .
\end{equation}
In the expanding Universe, where $n_e\sim 1/t^2$ 
the free path for photon
increases in time and $R \sim t^{3/2}$. 
But in any case $R$ stays always
less than the radius of the horizon, $L$.
The upper limit of $ R$ is $R_0$ 
at the moment of the hydrogen recombination.
At this moment the electron density becomes so small 
that photons after the
last scattering become free. 
Therefore $R_0$ is the last free path for photons
before they become free. The simple numerical estimates give
$R_0 \approx (0.15 - 0.2)L_0$, with 
$L_0$ being the horizon size at $z=1100$.
This value of $R_0$ corresponds to the 
angular size $\theta_0 \approx 20'$.
It is easy to show that this result does not actually depend 
on the initial
radius $R_i$ and $z_i$ of the CPS
 (for $z_i \gg 10^3$).

{\bf 3}.  It is very important that the intensity 
of radiation $I$ of these bubbles does not
depend, in fact, on detail of energy transfer and transport. 
Indeed, if the
radiation of the CPS may convert 
in several forms such as heating of matter, new
electron-positron pairs, shocks or any hydrodynamic motions,
then {\it always} the
total energy within the bubble 
is keeping {\it constant}. 

Then, the expansion of the Universe leads to the
dissipation of hydrodynamic motions, 
annihilation of electron-positron pairs
and total thermalisation of matter and radiation within 
the bubble with radius $R_f\le L$. 
This final radius $R_f$ is formed at the moment really not less
than $z=10^7$.
It is obvious that all energy emitted by this object will 
be trapped within
this radius. 
The only part of the total energy which is transfered into
neutrino will escape the bubble.
But due to the thermodynamic equilibrium
condition, this part cannot be more than 10-20\%.

So, we may conclude, that approximately all energy 
will be in photons,
and all variants of energy transformations 
do not influence $R_0$
and $I_0$ -- the luminosity of bubble at $z=1100$.

{\bf 4}.  Due to the effect of redshift, 
the value of $I_0$, in contrast to $R_0$,
depends on $z_i$ under any other equal conditions. 
The spatial distribution
of $I$ within the bubble caused by diffusion is
\begin{equation}
 I = I_0 \exp(-r^2/2R_0^2)\: ,
\end{equation}
where $r$ is a distance from the center of the bubble. 
Such a distribution is independent
on the variations in time of intensity 
of primordial point source.
But $I_0$ is the function of time corresponding 
to the time variations of initial $I_i$.

{\bf 5}. The redshifts indicated in the tittle are not 
specific for this effect. 
In fact we can consider all other intervals
$10^{10} \ge z \ge 10^4$. 
However, there appears to be a new effect at smaller
redshifts. It is the effect of a non-total energy thermalisation.
Any spectrum of photons emitted in the interval of redshift
$10^8 \ge z \ge 10^5$ will be converted to the so called 
$Y$-distortions (see
Sunyaev i\& Zel'dovich 1970; 
Illarionov \& Sunyaev 1974a, 1974b). 
The energy spectrum
of objects irradiated after $z\approx 10^5$ 
remains approximately without changes.

{\bf 6}.  Another aspect of this problem is the possibility 
of consideration of the
evolution of energy distribution from extended sources. 
Really we can
use all previous reasons to describe all extended objects 
which have sizes
smaller than the current value of $L$. 
It means that we can consider a small scale
fluctuations of matter distribution from the tail of 
the primordial spectrum,
caused by inflation. 
Nowadays there is a wide spread opinion that these fluctuations
disappear due to viscosity and their energy are 
totally redistributed in
space (see Silk 1968). So one cannot get any 
information about this part of
the spectrum directly from observations.

What is really happened ? 
All small wavelength sound waves are really
damped, but all their energy are not transported 
far from the initial region.
It is transformed from the kinetic into the thermal 
energy of matter (the smallest
part of it) and mainly in the energy of the CMBR. 
The final size of the bubble where
this energy will be distributed is $R_0$.

So on the base of our new result we can say, 
that actually all small scale
fluctuations after dissipation will 
influence the spectrum of
angular scale fluctuations 
on the scale size $ \theta_0 \approx 20'$. The
amplitude of this contribution may depend 
on the frequency of photons (see point 5).
If statistic of these fluctuations is taken into account,
one can get some
changes in the angular spectrum at the scales 
$\theta \ge \theta_0 $ with relatively lower amplitude.

{\bf 7}.  In conclusion we would like to emphasize
that the correct consideration of the process of
energy transfer and transportation from a point 
like source or a small extended object in highly dense
medium leads to a very slow expansion of region 
where all energy is
concentrated. 

From this it follows that the observations of the angular
fluctuations of the CMBR at the angular scale
$\theta_0 \simeq 20'$ will show up the presence of the
primordial CPSs, their statistics and their physical
properties. Besides one can obtain the information
about the small scale density fluctuations caused by inflation.

\bigskip\noindent

\begin{center}
{\large References}
\end{center}

\bigskip\noindent
Illarionov A. F., \& Sunyaev R. A., SvA, {\bf 51}, 698 (1974a)

\bigskip\noindent
Illarionov A. F., \& Sunyaev R. A., SvA,  {\bf 51}, 1162 (1974b)

\bigskip\noindent
Silk J., Ap.J., {\bf 151}, 459 (1968)

\bigskip\noindent
Sunyaev R. A., \& Zel'dovich Ya. B.,  Astrophys. Space Sci., 
{\bf 7}, 3 (1970)

\end{document}